# Is User Perception the Key to Unlocking the Full Potential of Business Process Management Systems (BPMS)?

## Enhancing BPMS Efficacy Through User Perception

Alicia Maria Martín Navarro
https://orcid.org/0000-0002-9443-6491
*INDESS, University of Cadiz, Spain*

María Paula Lechuga Sancho
https://orcid.org/0000-0003-2340-7615
*INDESS, University of Cadiz, Spain*

Marek Szelągowski
https://orcid.org/0000-0002-5114-6793
*Systems Research Institute Polish Academy of Sciences, Poland*

Jose Aurelio Medina-Garrido
https://orcid.org/0000-0003-3120-6426
*INDESS, University of Cadiz, Spain*

## ABSTRACT

This study investigates factors influencing employees' perceptions of the usefulness of Business Process Management Systems (BPMS) in commercial settings. It explores the roles of system dependency, system quality, and the quality of information and knowledge in the adoption and use of BPMS. Data were collected using a structured questionnaire from end-users in various firms and analyzed with Partial Least Squares (PLS). The survey evaluated perceptions of service quality, input quality, system attributes, and overall system quality. The findings indicate that service quality, input quality, and specific system attributes significantly influence perceived system quality, while system dependency and information quality are predictors of perceived usefulness. The results highlight the importance of user training, support, and high-quality information in enhancing satisfaction and BPMS. This research offers empirical evidence on the factors impacting user perceptions and acceptance, emphasizing the need for user-centric approaches in BPMS.

### KEYWORDS

Business Process Management Systems, BPMS, User Perception, System Quality, Information Quality, System Dependency

## INTRODUCTION

Michael Hammer (1996) argued that all organizations that want to function in the 21st century will have to focus on business process management (BPM). Although there is no total consensus, BPM is a methodology for evaluating, analyzing, and improving key processes based on the needs and desires of customers (Reijers, 2006). It represents a customer-oriented approach to managing processes within an organization. Scientific management (Taylor, 1911) considered the precursor of BPM, referred only to structured, repeatable mass production processes. However, over more than 100 years of development, BPM has expanded to encompass all organizational processes, regardless of their nature or place in the value chain (Mendling et al., 2020; Szelągowski, 2019). This has resulted in the practical need to adapt the method of identifying, modeling, implementing, and managing business processes, and, as a consequence, it has forced the evolution of information technology system supporting BPM (Szelągowski et al., 2022). They have evolved from traditional workflow systems, which automate business processes but lack flexibility and integration with other tools, into BPM systems (BPMS). BPMS offer greater flexibility, allow process modeling by the owner, and easily integrate with other technologies, representing a significant advancement over traditional workflows (Martín-Navarro et al., 2018). These systems support the automation of even unpredictable business processes. Additionally, with the advent of process mining techniques, it has become possible to obtain data on the implementation of these processes. This data facilitates artificial intelligence learning and the continuous improvement of the quality of support provided to users (Gartner, 2016; van der Aalst, 2013; van der Aalst et al., 2005). According to the principles of hyperautomation, BPMS allow for the flexible integration of external applications and system modifications without required programming (Gartner, 2019). BPMS are designed not to replace existing applications but to incorporate their use into new processes and leverage their information. This integration allows for more flexible and adaptable process modeling (Arjonilla Domínguez & Medina Garrido, 2009; Wong, 2013).

Undoubtedly, the use of the software affects the way work is carried out as well as employee competencies and human resource management (Dumas et al., 2023). However, over time, user perception of business processes has evolved. When processes were mostly manual, employees had direct control over each task, which initially led to resistance to automation, as systems were perceived as complex or even as a threat to their role within the company. As processes began to be automated, challenges arose, such as the need for continuous training to ensure that employees could adapt and fully leverage the potential of BPMS. Over time, perceptions evolved, and users began to recognize the benefits of reducing time spent on repetitive, manual tasks, allowing for a more strategic focus on their roles. Today, users emphasize significant improvements in data quality and an increased ability to manage complex processes more efficiently. Thus, the way users perceive the usefulness of the tool has become more critical than ever, as the effectiveness of BPMS usage largely depends on how well users adopt and integrate these tools into their daily work (Almatrodi et al., 2023).

Despite all of the above, academic research has yet to establish a consensus on the perceived utility of the systems' use by employees (Jalali, 2023). End-user acceptance is critical to the success of BPMS implementation. Without this acceptance, it is difficult to discuss the day-to-day use of these systems and even more difficult to use their support for innovative and creative implementation of business processes. Therefore, user feedback is considered a good indicator of IS implementation success (Asmah, 2016; Chang et al., 2015).

Similarly, most studies on BPMS use that acknowledge the end user's importance as a fundamental component of the process are primarily qualitative (Martín-Navarro et al., 2020). This provides fragmented knowledge about BPMS use (Poelmans et al., 2013). Given the growing importance and widespread adoption of BPMS, along with their role in enhancing process efficiency, this article aims to analyze these tools in a quantitative way. Specifically, it explores how various factors influence employees' perceived usefulness of BPMS. These factors include system dependency and the quality of both the information and the system itself. In response to Poelmans et al. (2013) calling for further research on these systems in commercial companies, we conducted a study using the partial least squares (PLS) methodology within this organizational context.

To achieve our research goal, this paper is structured as follows. First, we review the relevant theoretical background. Then, we detail the proposed model, define the variables, and outline the main hypotheses. Following that, we describe the research design and define the sample. We proceed to discuss the main results and highlight the study's conclusions, along with theoretical and practical implications for academics and practitioners. Finally, we address the study's limitations and outline future research directions.

## PRIOR RESEARCH AND HYPOTHESIS DEVELOPMENT

After reviewing 180 articles, DeLone et al. (1992) identified several methods for measuring the success of information systems. Information could be measured at different levels: the technical level, the semantic level, and the effectiveness level. The technical level was defined as the accuracy and efficiency of the system producing the information, the semantic level as the success of the information in conveying the intended meaning, and the effectiveness level as the impact of the information on the receiver (Shanon & Weaver, 1949).

The information system success model (ISSM), developed by DeLone et al. (1992), is one of the most widely used frameworks in academic literature for assessing the success of information systems. This model identifies six key dimensions that influence the success of these systems: system quality, information quality, system use, user satisfaction, individual impact, and organizational impact. According to the ISSM, system quality and information quality are direct determinants of system use and user satisfaction, while system use can positively or negatively affect user satisfaction and vice versa. The two factors, system use and user satisfaction, precede individual impact, which collectively affects organizational performance (DeLone et al., 1992).

This model has been widely adopted and adapted by numerous authors in academic literature to measure information system success based on user perceptions. Recent studies have continued to use the ISSM as a foundation for evaluating the perceived success of information systems, adjusting it to different contexts and types of technologies (Alyoussef, 2023; Martono et al., 2020; Sayaf, 2023; Wu et al., 2023).

Building on the ISSM, Poelmans et al. (2013) conducted a study on the operational success of BPMS. They defined these applications as ones that facilitate specific workflows and support various end users involved in executing these processes. To achieve their research objectives, they proposed a theoretical model describing perceived usefulness within BPMS. At the same time, they emphasized the need for a comprehensive success model for BPM applications, considering specific aspects: BPMS possess unique features supporting work coordination, such as task routing and assignment, which are crucial for success; the use of BPM applications is mandatory due to automated system records, making it difficult for users to avoid using the systems without detection; and integrating knowledge alongside data flow within processes automated by BPMS is essential. This knowledge is crucial for task execution within processes and is incorporated into distributed applications used by users—for example, mobile or web applications.

The operational BPMS research model by Poelmans et al. (2013) is considered the most suitable framework for evaluating information systems and their characteristics. This model is multifaceted and considers various attitudinal dimensions. We adopted the extended model by Poelmans et al. (2013) to define our research model, as seen in Figure 1. This approach helps us better understand users' perceptions and attitudes towards BPMS. By integrating an attitudinal dimension such as perceived usefulness, our approach allows for a more comprehensive understanding of users' perceptions and attitudes toward BPMS. This enhanced framework provides a thorough view of the effectiveness and impact of these systems from the end-user perspective. Thus, our BPMS research model combines technical and attitudinal dimensions to better understand the interaction between users and these systems, offering a solid foundation for evaluating and improving BPMS use in diverse organizational settings.

**Figure 1. Research model**

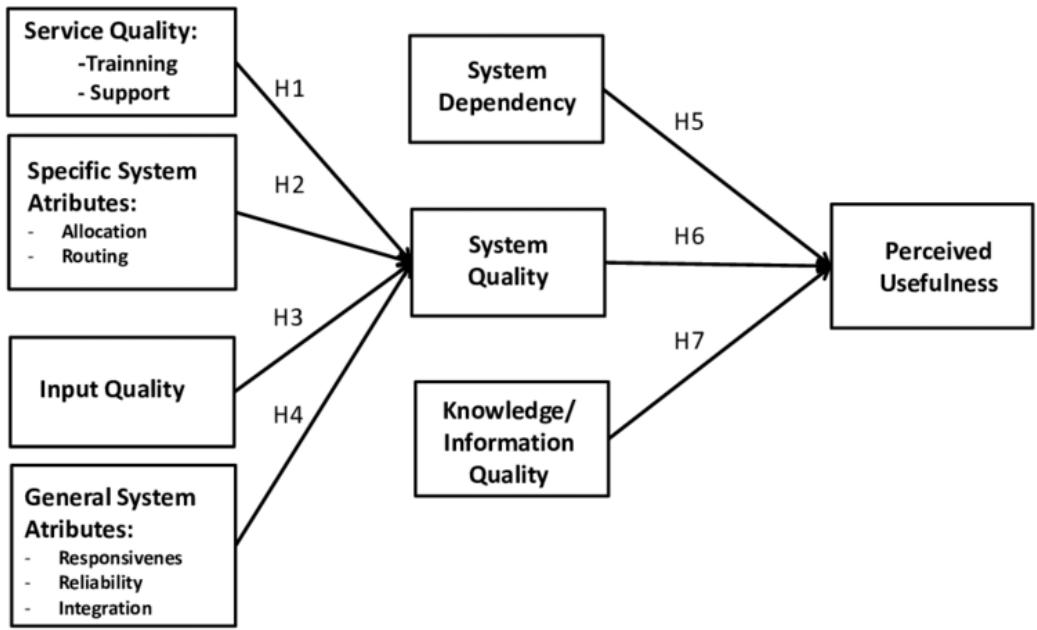

Source: Adapted from Poelmans et al. (2013)

The proposed model suggests that perceived usefulness is a direct consequence of system dependency, system quality, and information and knowledge quality. In turn, system quality is determined by service quality, input quality, and specific and general system attributes. However, it is important to note that while there are not many models on BPMS in the academic literature, the current models predominantly focus on the implementation phase (Juwitasary et al., 2018) and its impact on organizational performance (Ahmad, 2019). In contrast, our model stands out by focusing on the use of mature BPMS that have been in operation for several years. Furthermore, our model measures the success of the system based on perceived usefulness from the users' perspective, rather than solely on organizational outcomes.

## Service Quality

Service quality in IS has been extensively studied from various angles (Feng et al., 2014). It involves treating the IS user as an internal client and assessing the service quality by comparing user expectations with the actual service received. The quality of support provided to IS users, whether by an IT support department or other responsible personnel, determines service quality (Al-Okaily et al., 2023; DeLone & McLean, 2016; Ramírez-Correa et al., 2015). Service quality is perceived when users feel adequately attended to by specialists and when they have been properly trained to use the system. Employee training is essential for BPMS success (Hyötyläinen, 2015; Martín-Navarro et al., 2023). In this study, service quality is assessed based on user training and support from specialists to ensure correct interaction with the tool in daily tasks (Jo & Bang, 2023; Poelmans et al., 2013).

## Input Quality

Input quality refers to how well a BPM application enables end users to input data completely, understandably, sufficiently, relevantly, correctly, and in a timely way (Martín-Navarro et al., 2023).

This variable has not been widely used in IS success models in academic literature. Poelmans et al. (2013) found that input quality was a dimension previously unused in quantitative research. However, an earlier case study focused on data quality, highlighting the importance of designing a correct data input process within the BPMS architecture (Ofner et al., 2012). This underscores the relevance of this variable in the study. Considering that an information system involves inputs, transformations, and outputs (Campbell & Fogarty, 2021), assessing input quality appears crucial for system success.

## Specific System Attributes

One of the fundamental characteristics of BPMS is that they support the coordination of work. Among the aspects that the literature has investigated as improvement mechanisms for this coordination (Casati et al., 1998; Cugola, 1998) are routing and task allocation (Poelmans et al., 2013). Routing quality measures the degree to which an end-user encounters a work item; it consists of having the ability to send an activity back and forth through a process. On the other hand, assignment quality aims to examine how users rate the system's selection of tasks and work documents within each stage of the business process (Reijers & Poelmans, 2007). Specific system attributes is a construct specific to BPMS systems, as it represents characteristics that are specific to this type of tool and are not common in other types of software. Thus, it can be said that task assignment and adherence to a defined route are specific characteristics of BPMS (Poelmans et al., 2013).

## General System Attributes

Different dimensions or general attributes for measuring the quality of an information system can be found in the literature. Delone and McLean (2003) included: system reliability, integration with other tools, system functionalities, user responsiveness, and ease of use. Other authors identified dimensions such as: reliability, flexibility, integration, accessibility, and response time (Wixom & Todd, 2005). In this paper, general determinants of BPMS quality will be considered: reliability, integration, and responsiveness. Reliability refers to the user's dependence on the operation of the system, response time refers to the degree to which the system delivers the required results on time, and integration refers to how the system allows data from different sources to be integrated (Wu & Zhang, 2014).

## System Quality

System quality refers, in general, to the performance of the information system. It can be determined by its technical quality and by the components that are essential for processing, storing, and retrieving data (Ramírez-Correa et al., 2015). Petter et al. (2008) define system quality as those desirable characteristics of an information system. These include ease of use, flexibility, reliability, or response time. According to Wixom and Todd (2005), system quality represents a variable that measures the accessibility and flexibility of the system, taking into account the ease of use and availability of the BPMS. Similarly, Yusof and Yusuff (2013) consider a system to be of quality if the data it generates is accurate, if it is easy to learn and use, if it integrates with other applications, and if the speed of response is adequate. Ultimately, the quality of the system is a measure of the system's performance from an operational point of view.

From the above dimensions, our proposal relates service quality as a specific antecedent of system quality. This is because the more a user is trained or educated, as well as supported by specialists during his/her activities, the more he/she understands the system (Beatrix, 2022). In this way, he/she is able to see the capabilities of the tool and the application possibilities it shows him/her. This enables the user to use a system to its full potential, which raises the user's perceptions of the quality of the system (Nuryanti et al., 2021). Various studies suggest that a high level of service quality, including technical support, timely maintenance, and adequate user assistance, significantly contributes to improving system quality. Petter et al. (2013) highlight that the interaction between service quality and system quality is fundamental to ensuring the reliability and adaptability of the system, which directly impacts its performance. Likewise, Wixom and Todd (2005) argue that in modern environments, such as enterprise and cloud systems, continuous support and constant service updates critically influence the system's functionality and performance.

The quality of routing and mapping are specific characteristics of BPMS, forming part of the specific attributes of the systems. Thus, when users perceive that the system is working properly, their perception of the tool's quality improves. BPMS provide unique features, particularly in work coordination, where effective routing and allocation are critical to the success of BPM applications (Poelmans et al., 2013). Routing quality ensures that tasks are correctly assigned within the process, while allocation quality reflects user satisfaction with how tasks are distributed across different stages of the business process (Reijers & Poelmans, 2007). These attributes contribute to greater transparency in process management, allowing for individualized resource allocation and enhanced operational efficiency. Moreover, integrating advanced optimization techniques and dedicated resource management components within BPMS further elevates the systems' quality, enabling not only effective process execution but also more efficient use of available resources (Ihde et al., 2022).

On the other hand, several studies highlight the strong connection between data quality and system quality. For instance, research demonstrates that accurate and timely input data significantly enhance system performance. More specifically, the study by Gorla et al. (2010) on Hong Kong companies shows that when inputs are of high quality, systems are better equipped to provide accurate and timely information, which, in turn, improves decision-making processes and boosts overall organizational efficiency. This emphasizes the critical role of data quality in ensuring the effectiveness of information systems. Additionally, Poelmans et al.'s (2013) research on BPMS end users in four projects involving communal and governmental business processes found a key insight. Input quality was identified as a major determinant of system quality. Furthermore, system quality itself acts as an antecedent of perceived usefulness. This emphasizes the role of data quality in system effectiveness as well as how users perceive and benefit from the system. Therefore, input quality influences the quality a user expects from a system (Campbell & Fogarty, 2021).

The dimensions of reliability, integration, and responsiveness, which are general attributes of the system, reflect the ease of use, so that the user, through these dimensions, will consider the system to be of quality (Poelmans et al., 2013). In this regard, Nurkasanah et al. (2024) found that the general system attributes positively influence system quality in their research on end users of workflow systems. Similarly, Martín-Navarro et al. (2023) found a positive relationship between general system attributes and system quality in their study of BPMS users in commercial enterprises in Spain and Latin America. Their research highlighted that the overall attributes of BPMS contribute to enhancing the perceived quality of the system among end users. Based on the above, in relation to system quality, the following hypotheses are proposed:

**H1**: Service quality is positively related to system quality.
**H2**: Specific system attributes are positively related to system quality.
**H3**: Input quality is positively related to system quality.
**H4**: General system attributes are positively related to system quality.

## System Dependency

Software dependency represents a variable that, for Atadil et al. (2021), consists of the degree of interaction with the software that the user needs to perform the work for which he/she is responsible. Although the use of BPMS is considered mandatory (Poelmans et al., 2013), the intensity of this use is not always the same among employees. Different types of users can be distinguished, while some will use BPMS as their main tool, supporting many of their daily tasks; others will use it only occasionally, as a peripheral coordination tool to record or distribute some of their work output (Butt, 2020). Thus, the higher the dependence a user has on BPMS, the more important he/she will find them for his/her work tasks, regardless of the quality of the system. Conversely, when the user has little use for the system, he/she will hardly appreciate the value of the system (Poelmans et al., 2013). In essence, the system dependency determines the hours an end-user spends per week interacting with the BPMS he/she uses. This variable is observable and is measured through the questionnaire, in which the user specifically indicates the hours per week he/she spends with the BPMS.

## Information and Knowledge Quality

Information/knowledge quality can be defined as the degree to which the system generates information in a sufficient and appropriate way (Martín-Navarro et al., 2023; Sheng et al., 2020). Thus, it can be determined that information or knowledge quality refers to the desirable characteristics of the outputs of an information system and reflects, among other things, the accuracy, timeliness, completeness, relevance, consistency, and format of the information that is provided by the system. When information is generated appropriately, completely, on time, and accurately for a work task, perceptions of the usefulness of the system that supports those tasks will be higher (Poelmans et al., 2013; Ramírez-Correa et al., 2015). In the knowledge management environment, it is the user who distinguishes between information and knowledge. Whether a user distinguishes between information and knowledge will depend on when and what task he/she is performing (Holsapple, 2004). It is therefore common to refer to these two concepts interchangeably. When designing a system, the information it provides should be clear, understandable, and relevant to the task (Wu & Zhang, 2014). In one study, Wu and Zhang (2014) found a strong correlation between information quality and perceived usefulness in e-learning 2.0 systems and concluded that it would be of great interest for these systems to generate more relevant information for the user.

The information and knowledge quality is a variable that assesses the degree to which the system generates sufficient information and knowledge to be able to perform the tasks. This variable is composed of three aspects. These aspects comprise the assessment of the information provided by the system, the quality of the content, and the quality of the context and linkage (Poelmans et al., 2013; J.-H. Wu & Wang, 2006).

## Perceived Usefulness

The benefits of a system can be defined as the degree to which a user believes that using the application brings benefits to himself/herself and to the organization. These outcomes are often assumed to reflect an increase in job performance as well as an increase in productivity (Ramirez-Correa et al., 2015; Staples et al., 2002). These benefits can include enhancements such as market expansion, increased sales, or cost savings. However, quantifying these benefits objectively is often challenging due to environmental factors that influence measurements (McGill et al., 2003). Consequently, there is limited consensus on how to measure net benefits objectively, with such assessments often relying on the perceptions of those who use the system. As a result, perceived system benefits, also referred to as perceived utility, have been adopted as an important proxy measure of system success (Wixom & Watson, 2001). When users perceive a system as useful, they associate it with improvements in work performance, such as completing tasks more efficiently, accurately, and in less time (Al Shibly, 2014; Teo et al., 2008). This perception highlights the role of technology in enhancing task completion and overall effectiveness (Lateef & Keikhosrokiani, 2023).

Poelmans et al. (2013) found that reliance on BPMS is a positive determinant of perceived usefulness. On the other hand, Iivari (2005), in the field of medicine, measured the amount of daily use and frequency of system use. Thus, he found that the amount and frequency of system use had a strong impact on perceived usefulness. Similarly, in the work of Martín-Navarro et al. (2023), a positive relationship was found between system dependency and perceived usefulness, both for business processes and knowledge management among BPMS users. Their study revealed that the more users rely on BPMS for critical tasks, the greater their perception of the systems' usefulness.

The relationship between system quality and perceived usefulness has been studied by several authors in previous works. Although Jo and Park (2023) did not find this relationship to be significant, Ohanu et al. (2022) did find this in their study on the adoption of learning tools among university students. Similarly, in the work of Asmah (2016), this effect was found to be significant for users who utilized an electronic check clearing system. In this regard, Feng et al. (2014) examined website quality and discovered that page quality was positively correlated with perceived usefulness. The work of Wang and Yang (2016) on knowledge management systems (KMS) in small and medium-sized enterprises also found a relationship between system quality and perceived usefulness.

Wu and Wang (2006) found a highly significant positive influence of knowledge or information quality on perceived usefulness in KMS within the top 500 companies in Taiwan. Other authors such as Al Shibly (2014), in a study involving 200 users in Jordan, found that information quality is positively related to perceived usefulness in their research on electronic document management systems. Similarly, Feng et al. (2014) and Wu and Zhang (2014) have empirically demonstrated that information quality positively influences users' perceived usefulness. Feng et al. (2014) focused on online trading websites for securities brokers, while Wu and Zhang (2014) conducted their study with a sample of 284 participants from companies in China that have already implemented e-learn. In turn, Poelmans et al. (2013) also found significant relationships between information quality and perceived usefulness in their research on BPMS users in public administration. Moreover, according to the expectation-confirmation theory (Oliver, 1980), users form perceptions based on their experiences with the system. High-quality information exceeds users' expectations, thereby enhancing their perception of the system's usefulness. Therefore, we propose the following hypotheses:

**H5**: System dependency is positively related to perceived usefulness.
**H6**: System quality is positively related to perceived usefulness.
**H7**: Information/knowledge quality is positively related to perceived usefulness.

## METHOD

### Questionnaire Design

The data for the study variables were obtained through a questionnaire addressed to BPMS end users belonging to commercial companies. The questionnaire was developed based on existing surveys from the information systems literature, specifically focusing on BPMS (Poelmans et al., 2013) and KMS (Wu & Wang, 2006). It was tailored to assess the experiences of end users of BPMS, following the established theoretical framework. After the initial design, the questionnaire was reviewed by a panel of experts, including researchers and industry professionals, to check its content and formal quality. The panel included three members of a BPM software development company's training team and two modelers who create workflows using BPMS for companies with this software. These professionals were consulted through informal interviews, which led to improvements in the questionnaire's clarity for end users. Throughout this process, a critical analysis of the structure and content of the questionnaire was conducted, ensuring its alignment with the research objectives. The feedback from experts was synthesized to refine the instrument, providing clarity and precision in its questions. The final questionnaire was pre-tested by a pilot group of 17 end users of the tool. The users were carefully selected based on their experience with mature BPMS implementations. They provided valuable feedback on the clarity and ease of answering the questions. Based on their input, specific adjustments were made to refine the wording and structure of the questions, ensuring they were comprehensible and effectively captured the intended information. The pre-test results were discussed, and a final synthesis of the necessary improvements was integrated into the final version of the questionnaire. This iterative process of analysis, discussion, and synthesis helped validate the questionnaire and enhance its reliability as a survey instrument.

## Data Collection

Once the debugging process concluded, following Lee et al. (2015), the final questionnaire was structured in two large, well-differentiated blocks. The first block includes the demographic variables. The second block of the questionnaire is based on a Likert scale, since this type of scale is easy to construct and apply. The scale is a seven-point Likert scale (Kwok, 2014; Lee et al., 2015) where one corresponds to "never" and seven to "always." Except for the variable system dependency, which is numerical and it is measured by the number of hours that the user spent in the system.

Data collection was carried out between November 2018 and February 2019 through a directed study. These types of studies have been conducted by many other authors who have conducted empirical, quantitative research in the area of information systems or other types of information and communication technology (Feng et al., 2014; Hariguna et al., 2016). This type of study is even more necessary when the implementation of the analyzed information system is very low among the population of companies.

To conduct data collection, a self-administered online questionnaire was created using the Google Forms tool, consistent with previous similar work (Hariguna et al., 2016; Jiang & Wu, 2016; Lian, 2017; Martín-Navarro et al., 2021). This type of method is very useful as it allows users to access the survey through a web link. In addition, the data obtained is dumped directly into a spreadsheet, which facilitates the subsequent statistical processing of the data.

## Sample

The sample organizations were identified through the websites of BPMS providers. Data were collected from a targeted sample of users within companies that had implemented BPMS. A total of 53 commercial companies that had been using BPMS for more than two years were contacted through the BPMS managers in each organization. Out of all of the companies contacted, only 12 agreed to participate in this study, all of which were located in Latin America or Spain. Participating companies in our study were sent 415 questionnaires to disseminate to employees using BPMS. To streamline dissemination, a standard letter was created for end users, briefly explaining the project and its objective. These companies generated a total of 242 surveys completed by end users of different BPMS, with a response rate of 58.31%. It was verified that there were no missing data. For the variable hours spent per week, the data corresponding to the interval "more than 40 hours" was imputed by the value 40. In addition, it was found that no variable needed to be rescaled. Of the 242 effective respondents, 73.6% are Spanish, 36.4% are women, 55.8% are between 25 and 40 years old, 25.6% have between 101 and 500 employees in their company, and three out of four have a degree, diploma, or bachelor's degree, as seen in Table 1.

Table 1. Demographic data

| Items | Types | N | % |
|---|---|---|---|
| Experience with BPMS (in years) | 35.57 (2.9 years) | | |
| Dependence on BPMS (hours per week) | 10.29 | | |
| Sex | Man | 154 | 63.6 |
| | Woman | 88 | 36.4 |
| Age | Less than 25 | 6 | 2.5 |

**Table 1. Continued**

| Items | Types | N | % |
|---|---|---|---|
| | From 25 to 40 | 135 | 55.8 |
| | From 41 to 55 | 88 | 36.4 |
| | More than 55 | 13 | 5.4 |
| Educational level | Baccalaureate | 7 | 2.9 |
| | Degree/diploma/bachelor's degree | 182 | 75.2 |
| | Postgraduate | 38 | 15.7 |
| | Primary | 1 | 0.4 |
| | Secondary | 14 | 5.8 |
| Software | Aura Portal | 135 | 55.8 |
| | Bizagi | 16 | 6.6 |
| | PLAN Cibernos | 75 | 34.3 |
| | SAP | 8 | 3.3 |

## METHODOLOGY

To test the proposed hypotheses, a PLS structural equation model (SEM) was proposed following the methodological recommendations of Hair et al. (2014). Specifically, the SmartPLS 3.0 software (Ringle et al., 2015) was used to perform the analysis. The choice of this methodology and, in particular, the data collection process, the constructs investigated, and the measures used are suitable for the empirical analysis of relationships between related theoretical variables. In addition, the PLS-SEM approach allows for a comprehensive analysis of both the measurement and structural models, providing a robust framework for testing complex relationships. This methodological approach ensures that the analysis is robust and facilitates the subsequent synthesis and discussion of the results.

The model evaluation was conducted in three steps: global model assessment, measurement model assessment, and structural model assessment, even though the parameters of the measurement model and the structural model are estimated in a single step. Following the recommendations of Chin (2010) and Hair et al. (2014), the results were presented by first assessing the measurement model and then evaluating the significance of the model parameters. This approach allowed for rigorous analysis and ensures that the research has valid and reliable measures before drawing conclusions about the relationships between constructs.

## RESULTS

### Global Model

According to the recommendations of Henseler et al. (2016b), the overall goodness of fit of the model is the starting point of the model assessment. Although we use the PLS algorithm that does not require an analysis of the global model for the moment, the analysis was carried out with the fit measure standardized root mean square residual. This measure indicates a good fit of the model since its value (0.063) is below the maximum value of 0.08 imposed by Hu and Bentler (1998).

After checking the fit of the overall model, the measurement model was analyzed to ensure that the model had valid and reliable measures before we studied the relationships between the variables.

## Measurement Model

This section analyzes whether the theoretical concepts are correctly measured through the observed variables or indicators. Since all constructs are reflective, reliability is analyzed first. Subsequently, validity is analyzed (Henseler et al., 2016b).

Table 2 shows the individual reliability analysis for each item. While some indicators had loadings below the 0.707 threshold (Carmines & Zeller, 1979), their removal was not justified, as it would not significantly improve the reliability and validity of the remaining indicators. Furthermore, these indicators were retained due to their essential contribution to content validity (Hair et al., 2014). In any case, we do not have indicators with loadings below 0.4 that should be eliminated. Internal consistency, or construct reliability, was analyzed using composite reliability and Cronbach's alpha coefficient measures. Cronbach's alpha values exceeding 0.7 for all constructs indicate high reliability (Cronbach & Meehl, 1955; Streiner, 2003), and the composite reliability should be higher than 0.8 (Nunnally & Bernstein, 1995). The values of all constructs for both measures were found to be above 0.8. Once the reliability assumption had been analyzed, convergent validity, used to check that a set of indicators represents a single underlying construct, and discriminant validity, used to analyze the extent to which a given construct is different from other constructs, were examined. Convergent validity was demonstrated using average variance extracted (Henseler et al., 2009), and each construct was found to explain at least 50% of the variance of the assigned indicators, as seen in Table 2.

**Table 2. Evaluation of the measurement model under the complete theoretical model proposed and the parameters for the validation of the measurement model**

| Construct/Indicator | VIF | Weight | Loads | $\rho A$ | Cronbach's α | Average variance extracted |
|---|---|---|---|---|---|---|
| **Service quality** | 1.831 | | | 0.948 | 0.946 | 0.861 |
| CS1. The formation/training that I received was good | | 0.252 | 0.905 | | | |
| CS2. In general, I received sufficient training to be able to work with the system | | 0.264 | 0.925 | | | |
| CS3. In general, I'm being supported to be able to work properly with the system | | 0.283 | 0.943 | | | |
| CS4. I receive sufficient support to work with the system | | 0.279 | 0.938 | | | |
| **General system attributes** | 2.619 | | | 0.900 | 0.895 | 0.617 |
| AGS1. The system is available when I require it. | | 0.160 | 0.797 | | | |
| AGS2. The information that I use remains in the system (it does not get lost). | | 0.158 | 0.677 | | | |
| AGS3. The system works correctly (it does not get stuck). | | 0.178 | 0.757 | | | |
| AGS4. The reaction time of the system is correct. | | 0.203 | 0.882 | | | |
| AGS5. The speed of the system is sufficient for my purposes. | | 0.194 | 0.868 | | | |
| AGS6. can use the system in combination with other tools (word, excel, e-mail..) | | 0.192 | 0.769 | | | |
| AGS7. Tools, such as word, excel, e-mail. . ., are well integrated into the system. | | 0.185 | 0.728 | | | |
| **Specific system attributes** | 3.659 | | | 0.879 | 0.863 | 0.713 |
| AES1. The system allows selecting files/work items from the activity received to be able to perform it. | | 0.337 | 0.906 | | | |

**Table 2. Continued**

| Construct/Indicator | VIF | Weight | Loads | ρA | Cronbach's α | Average variance extracted |
|---|---|---|---|---|---|---|
| AES2. The system (re)distribute files/work items among your colleagues with the same role. | | 0.260 | 0.804 | | | |
| AES3. The system forwards work items to the next step/activity. | | 0.317 | 0.906 | | | |
| AES4. The system can put work items back into previous steps. | | 0.265 | 0.750 | | | |
| **Input quality** | 3.573 | | | 0.948 | 0.946 | 0.787 |
| CE1. I have sufficient data entry facilities in the system. | | 0.178 | 0.880 | | | |
| CE2. I can insert the data in a clear and understandable way. | | 0.198 | 0.912 | | | |
| CE3. I have sufficient means to correct and/or change the data in the system. | | 0.176 | 0.831 | | | |
| CE4. I have sufficient help/support when inserting data. | | 0.204 | 0.916 | | | |
| CE5. I can enter data when you need to enter data in the system. | | 0.185 | 0.873 | | | |
| CE6. I can enter the data in sufficiently detailed way. | | 0.186 | 0.907 | | | |
| **System quality** | 3.424 | | | 0.928 | 0.924 | 0.771 |
| CSI1. The system was easy to learn. | | 0.204 | 0.865 | | | |
| CSI2. The system is easy to use. | | 0.231 | 0.926 | | | |
| CSI3. The system does what I want it to do (without too much effort). | | 0.252 | 0.916 | | | |
| CSI4. The system is user friendly. | | 0.230 | 0.913 | | | |
| CSI5. The system is stable (It is tested and does not produce errors). | | 0.222 | 0.758 | | | |
| **Information/Knowledge quality** | 3.410 | | | 0.977 | 0.974 | 0.675 |
| CIC1. The information is reliable and accurate. | | 0.061 | 0.820 | | | |
| CIC2. The information is complete. | | 0.064 | 0.840 | | | |
| CIC3. The information is readable and easy to understand on the screen. | | 0.063 | 0.849 | | | |
| CIC4. Electronic presentation/format of the information (on the screen) is adequate. | | 0.061 | 0.837 | | | |
| CIC5. Printed version/presentation of the information is adequate. | | 0.059 | 0.780 | | | |
| CIC6. The speed with which the information can be gathered/retrieved is adequate. | | 0.060 | 0.794 | | | |
| CIC7. The information is updated in the system. | | 0.058 | 0.789 | | | |
| CIC8. The available information in the system is sufficient for my tasks. | | 0.066 | 0.881 | | | |
| CIC9. I have sufficient access to the information available in the system. | | 0.065 | 0.874 | | | |
| CIC10. The system makes it easy for me to create knowledge documents. | | 0.060 | 0.835 | | | |
| CIC11. The words and phrases in contents provided by the system are consistent. | | 0.062 | 0.870 | | | |

**Table 2. Continued**

| Construct/Indicator | VIF | Weight | Loads | $\rho A$ | Cronbach's α | Average variance extracted |
|---|---|---|---|---|---|---|
| CIC12. The content representation provided by the system is logical and fit. | | 0.065 | 0.883 | | | |
| CIC13. The knowledge or information provided by the system is available at a time suitable for its use. | | 0.061 | 0.861 | | | |
| CIC14. The knowledge or information provided by the system is important and helpful for my work. | | 0.067 | 0.893 | | | |
| CIC15. The knowledge or information provided by the system is meaningful, understandable and practicable. | | 0.069 | 0.917 | | | |
| CIC16. The knowledge classification or index in the system is clear and unambiguous. | | 0.065 | 0.891 | | | |
| CIC17. The system provides contextual knowledge or information so that I can truly understand what is being accessed and easily apply it to work. | | 0.065 | 0.867 | | | |
| CIC18. The system provides complete knowledge portal so that I can link to knowledge or information sources for more detail inquire. | | 0.054 | 0.674 | | | |
| CIC19. The system provides accurate expert directory. | | 0.044 | 0.587 | | | |
| CIC20. The systemprovides helpful expert directory (link, yellow pages) formy work. | | 0.041 | 0.574 | | | |
| **System dependency** | 1.025 | | | 1.000 | 1.000 | 1.000 |
| DS1. Number of hours the user spends using the tool per week | | 1.000 | 1.000 | | | |
| **Perceived usefulness** | | | | 0.954 | 0.953 | 0.876 |
| UP1. The system is very well suited to do the tasks that it is supposed to do. | | 0.267 | 0.932 | | | |
| UP2. Using the system enables me to handle my cases/ work items well. | | 0.276 | 0.952 | | | |
| UP3. In using the system, I can do my tasks in the process more efficiently. | | 0.273 | 0.937 | | | |
| UP4. The system really has added value in the business process. | | 0.253 | 0.921 | | | |

Finally, the discriminant validity of each construct was analyzed, which is the ability of a model to distinguish between different constructs. Although classical methods such as cross-loading analysis and the criteria of Fornell and Larcker (1981) have traditionally been used, according to Henseler et al. (2016a), these have shortcomings; they specify that discriminant validity is best detected by means of the heterotrait-monotrait ratio (HTMT). Specifically, the bootstrap 90% confidence intervals of HTMT were used, and it was found that no interval includes the value one so there is discriminant validity (Henseler et al., 2015). Table 3 shows the results for discriminant validity, highlighting that the HTMT values are below one, indicating that the constructs measured are distinct from each other.

**Table 3. HTMT confidence intervals**

| Construct/Indicator | Original sample (O) | 5.0% | 90% |
|---|---|---|---|
| General system attributes -> specific system attributes | 0.834 | 0.758 | 0.898 |
| Input quality -> specific system attributes | 0.905 | 0.850 | 0.954 |
| Input quality -> general system attributes | 0.809 | 0.748 | 0.866 |
| Information/Knowledge quality -> specific system attributes | 0.854 | 0.789 | 0.913 |
| Information/Knowledge quality -> general system attributes | 0.802 | 0.750 | 0.849 |
| Information/Knowledge quality -> input quality | 0.888 | 0.854 | 0.918 |
| Service quality -> specific system attributes | 0.716 | 0.621 | 0.802 |
| Service quality -> general system attributes | 0.629 | 0.544 | 0.713 |
| Service quality -> input quality | 0.649 | 0.563 | 0.731 |
| Service quality -> information/knowledge quality | 0.631 | 0.535 | 0.723 |
| System quality -> specific system attributes | 0.785 | 0.694 | 0.864 |
| System quality -> general system attributes | 0.777 | 0.709 | 0.839 |
| System quality -> input quality | 0.830 | 0.777 | 0.879 |
| System quality -> information/knowledge quality | 0.884 | 0.841 | 0.920 |
| System quality -> service quality | 0.630 | 0.529 | 0.725 |
| System dependency -> specific system attributes | 0.181 | 0.076 | 0.281 |
| System dependency -> general system attributes | 0.179 | 0.094 | 0.275 |
| System dependency -> input quality | 0.131 | 0.053 | 0.239 |
| System dependence -> information/knowledge quality | 0.142 | 0.058 | 0.242 |
| System dependency -> system quality | 0.204 | 0.121 | 0.282 |
| System dependency -> system quality | 0.159 | 0.074 | 0.249 |
| Perceived usefulness -> specific system attributes | 0.815 | 0.740 | 0.881 |
| Perceived usefulness -> general system attributes | 0.708 | 0.631 | 0.776 |
| Perceived usefulness -> input quality | 0.808 | 0.751 | 0.860 |
| Perceived usefulness -> information/knowledge quality | 0.914 | 0.888 | 0.939 |
| Perceived usefulness -> service quality | 0.619 | 0.524 | 0.705 |
| Perceived usefulness -> system quality | 0.835 | 0.781 | 0.884 |
| Perceived usefulness -> system dependency | 0.169 | 0.060 | 0.270 |

Therefore, we can say that the model is reliable and valid, so the validity of the measurement model has been tested and we can analyze the structural model.

## Structural Model

Once the goodness of fit is assured, we proceed to assess the structural model, as seen in Table 4, by studying possible problems of collinearity, assessment of the coefficient of determination, evaluation of the algebraic sign, magnitude and statistical significance of the path coefficients, and the size of the effects of the relationships. Since the estimates of the path coefficients are made in Ordinary Least Squares (OLS) regressions, the presence of multicollinearity between variables must be avoided. The Variation Inflation Factor (VIF) values were analyzed, and it was found that for all construct pairs they are less than five, confirming the independence of predictor variables and indicating that there is no collinearity (Belsley, 1991; Hair et al., 2014).

A good measure to analyze predictive power is the coefficient of determination ($R^2$) for the latent dependent variables, and it indicates the amount of variance in the construct that is explained by the model. Chin (1998) establishes the following reference values: from 0.67 substantial; from 0.33 moderate; and from 0.19 weak. In our case we will use the adjusted $R^2$, which controls for the complexity of the model. A substantial rating was observed for the two endogenous variables, system quality and BMPS perceived usefulness, for processes with values of 0.648 and 0.784 respectively, indicating a high predictive power.

Third, the significance assessment of the path coefficients is obtained with the bootstrapping technique. The subsamples were created with 5,000 observations and a one-tailed student's t-distribution and using the bias-corrected and accelerated bootstrap type for the estimation of confidence intervals (Hair et al., 2014). According to Henseler et al. (2009), intervals that do not include the value zero mean that the relationship is significant.

In addition, direct effects were analyzed using Cohen's (1988) scale, which indicates that if an $f^2$ value is between 0.02 and 0.15, the effect is small, between 0.15 and 0.35, the effect is moderate, and greater than 0.35, the effect is large. Table 5 summarizes the effects on the endogenous constructs, meaning the dependent variables that the study aims to explain based on other independent variables. Most of the hypotheses have a significant effect on the endogenous variables, except for H2 and H5. This suggests that not all factors are equally important in determining system quality or perceived usefulness. The greatest direct effect is observed in the impact of information/knowledge quality on perceived usefulness, as observed in H7, indicating that this is a key factor in users' perception of BPMS' usefulness. The $R^2$ values indicate that system quality and perceived usefulness are well explained by the independent variables in the model, with an especially high value for perceived usefulness (0.784).

**Table 4. Summary of effects on endogenous constructs**

| Hip. | Effects on endogenous variables | Suggested address | Direct effects | t-value (bootstrapp) | Percentile 95% confidence interval | Sign. | Explained variance (%) | Size of effects |
|---|---|---|---|---|---|---|---|---|
| | *System quality ($R^2$ = 0.648/ $Q^2$ = 0.460)* | | | | | | | |
| (H1) | Service quality -> System quality | + | 0.113 | 1.770 | (0.009; 0.216) | Yes | 6.68 | 0.020 |
| (H2) | specific system Attributes -> system quality | + | 0.070 | 0.662 | (-0.107; 0.239) | No | 4.91 | 0.004 |
| (H3) | input quality -> System quality | + | 0.475 | 5.082 | (0.310; 0.618) | Yes | 36.94 | 0.182 |
| (H4) | general system Attributes -> system quality | + | 0.237 | 2.925 | (0.105; 0.373) | Yes | 16.81 | 0.062 |
| | *perceived usefulness ($R^2$ = 0.784/ $Q^2$ = 0.642)* | | | | | | | |
| (H5) | System dependency -> perceived usefulness | + | 0.034 | 1.100 | (-0.022; 0.079) | No | 0.56 | 0.005 |
| (H6) | System quality -> perceived usefulness | + | 0.152 | 2.312 | (0.048; 0.262) | Yes | 11.97 | 0.032 |
| (H7) | Information/Knowledge quality -> perceived usefulness | + | 0.749 | 12.525 | (0.644; 0.841) | Yes | 66.07 | 0.771 |

*Note.* The variance explained represents the variance that each predictor explains for each endogenous variable. The percentage is estimated as the product of the correlation between the predictor with the endogenous and the value of the direct effect.

In addition, indirect effects were analyzed, and it was found that there is a significant relationship between general system attributes and input quality with respect to BPMS' perceived usefulness, as mediated by the quality of the system. These results are reflected in Table 5.

Table 5. Significance of indirect effects

| Indirect effects | Path coefficients | T-value (bootstrapp) | P value | Percentile 95% confidence interval | Significance |
|---|---|---|---|---|---|
| Specific system attributes -> Perceived usefulness | 0.011 | 0.596 | 0.276 | (-0.011; 0.049) | No |
| General system attributes -> Perceived usefulness | 0.036 | 1.930 | 0.027 | (0.013; 0.078) | Yes |
| Input quality -> perceived usefulness | 0.072 | 2.061 | 0.020 | (0.025; 0.141) | Yes |
| Service quality -> perceived usefulness | 0.017 | 1.232 | 0.109 | (0.002; 0.049) | No |

The first four hypotheses are those that refer to the characteristics of BPMS. Thus, service quality, seen from the training and technical support that a user receives, has a positive effect on the quality of the system. This positive effect has also been found for general system attributes with reference to system quality. In this way, users consider BPMS to be a business process automation technology with high reliability and great responsiveness. The positive relationship between input quality and system quality is noteworthy. This finding supports the results of the study by Poelmans et al. (2013), which found the effect of input quality on system quality to be highly significant, although in this case the relationship is rather moderate. However, the present study contrasts with Poelmans et al.'s (2013) in terms of the relationship between specific system attributes and system quality, as the hypothesis regarding this connection has been rejected in our findings. This may be because, to a user of BPMS, the assignment of tasks and the routing of tasks are transparent. That is, a user does not detect these tasks performed by the system; he/she simply receives and sends his/her own tasks but does not detect the routing of the process. This suggests that while BPMS users recognize the importance of input quality, general system attributes, and service quality, the overall contribution of specific system attributes to system quality may be perceived differently.

On the other hand, the findings of Poelmans et al. (2013) on the positive effect of system dependency on perceived usefulness are not confirmed. Similarly, the evidence reported by Iivari (2005) on the relationship between system use, measured by amount of use and frequency, and perceived usefulness is also not supported.

In the present study, there is evidence that system dependence does not have a positive effect on perceived usefulness. This could be because frequent users become more critical of system shortcomings over time. Another reason might be related to the time spent using the system, as observed in Nurkasanah et al.'s (2024) study, where the hypothesis was also rejected with the justification that limited use of the tool leads to low dependency on the system. In our study, the average usage time was 10.29 hours per week, indicating that many users do not regularly engage with the BPMS, which in turn prevents them from perceiving its usefulness. Therefore, we cannot agree with this relationship, which was first introduced in a quantitative analysis by Poelmans et al. (2013) in reference to BPMS. Therefore, we cannot determine that the more BPMS are used, the more usefulness is perceived by the surveyed users.

Hypothesis 6, after analysis of the data, is found to be confirmed. Thus, system quality has a positive effect on perceived usefulness; therefore, we agree with the contributions of many other authors, including Poelmans et al. (2013) and Wang and Yang (2016), on KMS in small and medium-sized enterprises. This relationship was not direct, but it was concluded that both dimensions were related through another intermediate dimension, or mediating variable, which is the purpose of the system. This was not the case in the work on business intelligence systems by Mudzana and Maharaj (2015). In this work, using the same model as Wang and Yang (2016), the use of the system did not receive enough support to mediate between the quality of the system and the net benefits or utility. Also, Wu and Zhang (2014) found the relationship between system quality and perceived usefulness to be very weak in their study of e-learning systems. However, we cannot agree with Wu and Wang (2006), who, in their study on KMS, found that system quality impacts perceived usefulness. Ultimately, our research shows that the quality of the system is a factor that determines its usefulness in performing tasks. This may be explained by the importance users place on training and technical support, which are crucial for them to perceive BPMS as high-quality systems.

Wu and Zhang (2014) found a strong correlation between information quality and perceived usefulness in e-learning 2.0 systems and concluded that it would be of great interest for these systems to generate more user-relevant information. Our study similarly shows that the quality of information and knowledge has a positive influence on perceived usefulness, and it reinforces the findings of Poelmans et al. (2013) and Wu and Wang (2006). These authors found that this variable was a determinant of perceived usefulness. We also agree with Sultono et al. (2015) who conducted a study on academic information systems in a national university in Indonesia. We also corroborate the results of Okazaki et al. (2015) who successfully proved that the quality of information has a positive impact on perceived value. Other authors, such as Al Shibly (2014), Feng et al. (2014), Wu and Zhang (2014), and Wang and Yang (2016), have also empirically demonstrated that information quality positively influences the perceived usefulness of different information systems for a user. Indeed, many authors have noted the impact of this positive relationship. Specifically, we can say that it is understood that for the user, the information or knowledge generated by a system is valuable when using BPMS in their work. This outcome can be explained by the fact that when BPMS produce high-quality information and knowledge, users are more likely to find the systems useful, as they directly contribute to their ability to perform tasks effectively. All of the above are shown in Figure 2.

**Figure 2. Final model**

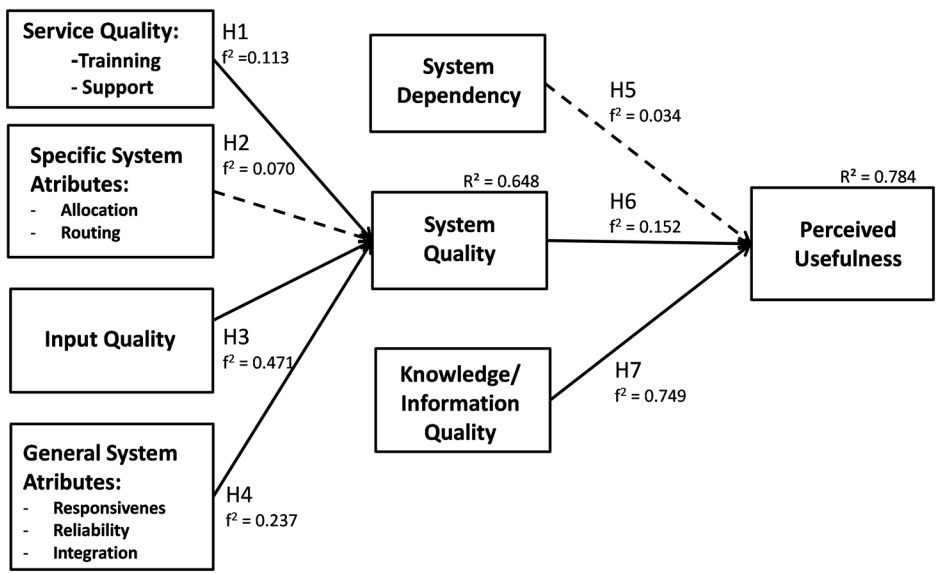

## DISCUSSION

Empirical data for the research by Poelmans et al. (2013) was collected through a field study among European users of four different BPMS applications during 2007 and 2008. The applications had been in operation for several years, so the use of these systems was considered stable (Poelmans et al., 2013). The data analysis from this empirical study demonstrated that information quality and system quality are antecedents to perceived usefulness from the user's perspective. These results partially align with a study conducted among engineers in Malaysia on the perceived usefulness of mobile KMS (Cheak et al., 2022). In that study, the authors found a significant relationship between information quality and perceived usefulness, which is consistent with our findings. However, unlike with our study, they did not find a significant relationship between system quality and perceived usefulness.

Furthermore, service quality is related to system quality, which aligns with the findings of Martín-Navarro et al. (2023) in their study on BPMS in commercial firms. Our results also indicated that input quality is the most important determinant of system quality, followed by application-specific characteristics, such as task allocation and routing. Specifically, task routing is more critical than task assignment quality. This suggests that the management of workflow through business processes is considered more important than the specific assignment of tasks to different participants (Poelmans et al., 2013).

The study revealed that users evaluate BPMS not only based on the quality of information generated by the system but also on how information is entered into the system. Additionally, the general attributes of BPMS, such as reliability, responsiveness, and integration with other tools, are important antecedents of system quality. These attributes are considered general because they are applicable to any information system.

### Theoretical Implications

Our empirical study generates a number of contributions to both academic research and the business sector. From an academic perspective, the quantitative approach provides empirical evidence that strengthens and expands the theory on the success of BPMS. Specifically, our work contributes to the existing knowledge on BPMS by offering results obtained from commercial companies, addressing a limitation highlighted by Poelmans et al. (2013). Additionally, it responds to Petter et al.'s (2008) call for more empirical studies on the dimensions of system success, demonstrating that service quality precedes system quality. Our findings show that input quality is also a key determinant of system quality in the BPMS context, an aspect that has been underexplored in the academic literature. In our study, we demonstrate that accurate, complete, and timely data entry into the system is essential for ensuring users' perceived quality. This suggests that in systems like BPMS, which depend heavily on complex process inputs, input quality plays a fundamental role in shaping the perceived effectiveness of the system. This finding introduces a new dimension to the ISSM by proposing that, in environments where data inputs are critical to business processes, input quality should be considered a direct antecedent of system quality. Moreover, we found that system dependency does not have a significant impact on perceived usefulness, suggesting that in highly automated environments like BPMS, continued use does not always lead to greater acceptance.

Our findings extend existing technology adoption models, such as the Technology Acceptance Model (TAM) and the Unified Theory of Acceptance and Use of Technology (UTAUT), by revealing that system quality perception and information quality are key predictors of perceived usefulness in BPMS contexts. These relationships, particularly the influence of input quality and facilitating conditions like technical support, play a critical role in shaping system acceptance, offering new dimensions for future adoption models, especially in automation-centered environments.

## Practical Implications

Regarding practical implications, this study provides valuable contributions to the professional world. Today, organizations recognize that incorporating technology is key to maintaining their competitiveness in the market (Konthong et al., 2016). To successfully implement these technologies, it is essential that employees accept and fully utilize their capabilities. The study concludes that BPMS are perceived as high quality when the service that accompanies them is also of high quality. To ensure this in practice, it is necessary to maintain high standards in three key areas: the integration of BPMS with the existing technological infrastructure; the implementation process and continuous support in their use; and data entry, management of business processes in the system, and the information or knowledge generated by the BPMS. Therefore, to fully harness the potential of BPMS, it is recommended to do the following:

1. From the BPMS selection phase, ensure they meet the requirements for integration with the current or future technological infrastructure, including systems and sensors (such as the internet of things) that may serve as data sources or recipients.
2. When preparing the implementation project, choose a methodology that aligns with the nature of the business processes and the organization's BPM maturity level. Additionally, it is crucial to provide employees with a training program suited to their competencies and offer them adequate support, especially in the early stages of use.
3. From the BPMS design phase and during the configuration of business process flows, create windows, dictionaries, and drop-down lists that facilitate data entry and retrieval as well as the management of the executed process flow.

The study has shown that these aspects are highly valued by users. Additionally, the BPMS must be reliable and always available to prevent information loss. In short, the systems must function properly. In addition, the systems must be fast in order to meet the user's purposes. And they will be highly valued if they can be integrated with other tools needed to perform the tasks.

Based on all of the above, if those in charge are able to follow these premises, those managing the implementation of BPMS will be able to achieve a higher-quality system, which will enhance users' perception of usefulness.

## LIMITATIONS AND FUTURE LINES OF RESEARCH

The results of the empirical research must be interpreted with some limitations in mind. Limitations are presented for the purpose of connecting them in the near future with new lines of research. Thus, the validity of the findings cannot be established on the basis of this single study, as the evidence is limited to only the 12 companies consulted. Therefore, the results should not be rashly overgeneralized. While the results offer valuable insights, their validity needs to be tested further with samples from different industries, organizational settings, and geographical contexts. Future research should consider the specific characteristics of the business processes being managed, such as their level of complexity and predictability and the degree of knowledge intensity required for their execution. Since most of the users surveyed are Spanish or Nicaraguan, and our sample predominantly comes from Spanish-speaking countries, it may limit the generalization of our findings to other cultural contexts. Therefore, it would be interesting to carry out the same study in different samples from other cultures or countries. Additionally, as commented by Wu and Wang (2006), there are other external factors that may influence the results obtained and that have not been taken into account in this research. These include, among others, the attitude of the management, the culture of the organization, and the confidence to answer the questionnaire freely.

In another sense, the novelty of the research topic and the fact that a robust theory on the success of BPMS has not been defined means that the theoretical model proposed as the basis for the study has developed its hypotheses based on Poelmans et al.'s (2013) study and other authors who have studied different types of information systems. It is therefore necessary to be able to contrast the results obtained with the findings of other BPMS research.

Future research should focus on strengthening the accepted hypotheses, which at the same time have confirmed the findings found in the literature. This would strengthen and build theoretical propositions and thus give more substance to the managerial theory of BPMS. For this, more empirical studies should be carried out in other samples and in other countries. Spain is the third European country, after Germany and England, that has been making progress for several years in the automation and management of processes. This makes the business fabric competitive both within and outside of Europe. Moreover, progress is being made not only in the field of private enterprise but also in public administration. For this reason, the study of the success of BPMS in the Spanish public administration is proposed as a future line of research.

On the other hand, the empirical analysis carried out and the limitations identified in the previous section open the door to interesting future lines of research. To begin with, future studies should pay closer attention to the growing influence of new technologies, such as robotic process automation and artificial intelligence, on BPMS systems. These technologies increasingly support users by automating data entry, managing process flows, and even assisting in decision-making, often relieving them from routine tasks. Their impact may significantly reduce the importance and labor intensity of simple task execution, shifting users' focus towards decision support within BPMS and identifying deviations from the established process flow.

Additionally, future research could explore the intersection between BPMS technology and legislative frameworks. While this study focuses on the technological and organizational aspects, there is a growing need to examine how BPMS can support managers in fulfilling their legal responsibilities (Perácek & Kaššaj, 2023) as well as align with telework regulations, particularly in light of the gaps in protection for teleworkers (Marica, 2022). This would provide a broader perspective on BPMS in ensuring compliance with both organizational and legal requirements. Finally, future quantitative research in the area of information systems should take into account the quality of the input as a determinant of the success of the systems.

## CONCLUSIONS

As seen in various papers on information systems, the most commonly used methodology for hypothesis testing is SEM. SEM is particularly interesting because it identifies causal relationships between latent and non-latent variables. Additionally, if the model does not fit the data, it allows for a model refitting process (Escobedo Portillo et al., 2016). The data collection process was challenging. Due to the lack of a comprehensive registry of companies using BPMS, the best way to access them was through the suppliers' websites. Professionals collaborating in the study distributed the online questionnaire to end users, resulting in 242 valid responses. Their collaboration was a significant step forward in the research.

The study found that five of the seven hypotheses proposed in the theoretical model were supported. Some findings were consistent with the literature review, while others were not. From a technical perspective, it was established that the quality of BPMS is determined by the quality of service, the quality of input, and the systems' general attributes. Users appreciate IT support when issues arise with the tools and value the training they receive. Regarding the quality of input, users believe that high-quality BPMS must provide convenient and useful data input. They also consider the systems' reliability and integration with other tools to be features that enhance quality. Since no empirical evidence was found for the relationship between specific BPMS attributes and BPMS quality, it can be argued that factors affecting system quality are necessary for any IT, not just BPMS. These factors include quality of service, quality of input, and general system attributes.

Regarding perceived usefulness for carrying out process-related activities, users consider the tool useful if it is of high quality and generates interesting information or knowledge. However, it does not appear that the more time a user spends with the tool impacts perceived usefulness.

## COMPETING INTERESTS STATEMENT

The authors of this publication declare there are no competing interests.

## FUNDING STATEMENT

This publication has been made possible thanks to funding provided by the Program for the Promotion and Encouragement of Research and Transfer of the University of Cadiz (Plan Propio UCA 2022-2023) and the University Research Institute for Sustainable Social Development (INDESS).

## PROCESS DATES

## CORRESPONDING AUTHOR

Correspondence should be addressed to María Paula Lechuga Sancho (Spain, paula.lechuga@uca.es)

*Dr. Alicia Martín-Navarro is an Associate Professor in Business Management Department, at Cádiz University, Spain. She holds a Business Administration Master's Degree and earned her PhD in Social Science. She has written some books about business decision-making. Her research interests are in the areas of entrepreneurship, information systems for management, and knowledge management. She has published in JCR journals and presented her research work at various international and national conferences, serving as a reviewer for some of them. She has been Visiting Professor at Universidade da Beira Interior in Covilhá (Portugal). Currently, she is assistant manager at INDESS (Research University Institute for Sustainable Social Development), University of Cadiz, Spain.*

*Dr. María Paula Lechuga is an Associate Professor in the Department of Business and Management at Cádiz University. She holds a Business Administration Master Degree, and she is currently participating in a national competitive project on bibliometric methods. Her work appeared in Expert Systems with Applications, Journal of Knowledge Management, Journal of Cleaner Production, Studies in Higher Education, Personnel Review, International Journal of Agile Systems and Management, Journal Industrial Management & Data System, Journal of Management and Enterprise Development, Spanish Accounting Review, International Journal of Management Education, Profesional de la Información, International Journal of Contemporary Hospitality Management, British Food Journal, among others. She has participated in more than 35 national and international conferences.*

*Dr. Marek Szelągowski is an experienced practitioner of business process management. Currently employed as assistant professor in Systems Research Institute of the Polish Academy of Sciences. For over 25 years, he has been involved in the implementation of IT solutions that support management based on common sense improvements and simplification of processes, selecting and implementing IT solutions appropriate to the client's situation. He participated in the development and implementation of IT solutions in the areas of accounting, human resource management, production, project management, IT infrastructure management, etc. Among other things, as the CIO of the Budimex Group he was responsible for the creation and development of the IT office, and most of all IT strategy for adapting to changing business needs. He is the author of numerous scientific articles in the field of process management and knowledge management as well as the monograph "Dynamic BPM in the Knowledge Economy: Creating Value from Intellectual Capital" devoted to the causes and consequences of dynamic BPM and the principles of its implementation in organizations, including the need to ensure effective knowledge management. In the AGH@BPM study, he was recognized as one of the five BPM researchers in CEE countries with the largest number of publications in reputable journals.*

*Dr. José Aurelio Medina-Garrido is a Professor of Management at Cádiz University (Spain). His research interests include human resource management, strategic management, and information systems management. His academic work has been published in several journals and books. He has papers published by International Entrepreneurship and Management Journal, International Small Business Journal, and International Journal of Hospitality Management, among others. He is also a business consultant.*